\documentstyle[12pt]{article}

\title{Decoherence Caused by Topology \\
in a Time-Machine Spacetime\footnote{Intern. J. Mod. Phys. D5, 1-27 
(1996)}}
\author{\large Michael B.Mensky$^{*\dag\ddag}$ and
Igor D.Novikov$^{\dag\ddag\P\copyright}$\\[10pt]
$^*$P.N.Lebedev Physical Institute, 117924Moscow, Russia\\
Email: mensky@sci.lebedev.ru\\[3pt]
$^{\dag}$NORDITA, Blegdamsvej 17\\
DK21-00 Copenhagen {\O}, Denmark\\[3pt]
$^{\ddag}$Theoretical Astrophysics Center, Blegdamsvej 17\\
DK-2100 Copenhagen {\O}, Denmark\\[3pt]
$^{\P}$Astro Space Center, P.N.Lebedev Physical Institute\\
Profsoyuznaya 84/32, Moscow 117810, Russia\\[3pt]
$^{\copyright}$University Observatory, {\O}ster Voldgade 3\\
DK-1350 Copenhagen K, Denmark
}

\date{}
\newtheorem{remark}{Remark}
\newcommand{\be}{\begin{equation}}
\newcommand{\ee}{\end{equation}}
\newcommand{\ba}{\begin{eqnarray}}
\newcommand{\ea}{\end{eqnarray}}
\newcommand{\ban}{\begin{eqnarray*}}
\newcommand{\ean}{\end{eqnarray*}}

\newcommand{\al}{\alpha}
\newcommand{\tr}{{\rm {Tr}}}

\begin{document}
\maketitle
\newpage
{Non-relativistic quantum theory of non-interacting particles
in the spacetime containing a region with closed time-like
curves (time-ma\-chine spacetime) is considered with the help
of the path-integral technique. It is argued that, in certain
conditions, a sort of superselection may exist
for evolution of a particle in such a
spacetime. All types of evolution are classified
by the number $n$ defined as the
number of times the particle returns back to its past. It
corresponds also to the topological class ${\cal P}_n$ of
trajectories of a particle. The evolutions corresponding to
different values of $n$ are non-coherent.
The amplitudes  corresponding to such
evolutions may not be superposed. Instead of one
evolution operator, as in the conventional (coherent)
description, evolution of the particle is described by a family
$U_n$ of partial evolution operators. This is done in analogy
with the formalism of quantum theory of measurements, but with
essential new features in the dischronal region (the region with
closed time-like curves) of the time-machine spacetime. Partial
evolution operators $U_n$ are equal to integrals $K_n$ over
the classes of paths ${\cal P}_n$ if the evolution
begins and ends in the chronal
regions. If the evolution begins or/and ends in the dischronal
region, the integral $K_n$ over the class ${\cal P}_n$ should be
multiplied by a certain projector to give the partial
evolution operator $U_n$. Thus defined partial evolution
operators possess the property of generalized unitarity $\sum_n
{U_n}^{\dagger}\,U_n = {\bf 1}$ and multiplicativity
$U_m(t'',t')\,U_n(t',t)=U_{m+n}(t'',t)$. In the last equation
however one of the numbers $m$ or $n$ (or both) must be equal to
zero. Therefore, the part of evolution containing repeated
returning backward in time cannot be factorized: all backward
passages of the particle have to be considered as a single act,
that cannot be presented as gradually, step by
step, passing through `causal loops'.  The (generalized)
multiplicativity and unitarity take place for arbitrary  time
intervals including i)~propagating in initial and final chronal
regions (containing no time-like closed curves) or from the
initial chronal region to the final one, and ii)~propagating
within the time machine (in the dischronal region),
from the time machine to the final
chronal region or from the initial chronal region to the time
machine.}

\vspace{0.5cm}
PACS number(s): 04.20Cv, 04.20.Ex, 04.60.Ds, 98.80.Hw.
\vspace{0.5cm}

\section{Introduction}\label{intro}

It is known that General Relativity predicts the possible
existence of spacetimes with non-trivial topology. The most
intriguing among them are spacetimes with a region (or regions)
containing closed time-like curves (CTC's). These {\em
time-machine spacetimes} will be considered in the present paper.
It is not clear whether the laws of physics permit the
development of CTC's in the course of the evolution of a
spacetime from some reasonable initial conditions (see
\cite{Th91}-\cite{Th93} and references therein). We will not
concern this problem here supposing that this is possible.
Instead of this, taking spacetime geometry {\em ad hoc}, we
shall investigate evolution of quantum particles in this
geometry.

We shall discuss only non-relativistic processes. In this case
the spacetime can be divided into {\em chronal regions} (regions
without CTC's) and {\em dischronal regions} (containing CTC's) by
slices $t=\mbox{const}$. For simplicity we shall discuss here the
spacetime with only one dischronal region. This dischronal region
(located between time moments $t_1$ and $t_2$) is preceded, at
$t<t_1$, and followed, at $t>t_2$, by chronal regions. We shall
call the first chronal region (at $t<t_1$) the initial region,
the second chronal region (at $t>t_2$) the final region, and the
dischronal region (at $t_1 < t < t_2$) the {\em time machine}.

In a series of rather recent papers (see for example
\cite{Fr90}-\cite{Th94}) the question was analyzed whether the
standard laws of physics (classical and quantum) can accommodate,
in a reasonable manner, a spacetime with CTC's. In this paper
we shall focus our attention on the problem of propagating
non-relativistic non-interacting quantum particles in the
spacetime with CTC's (the time-machine spacetime).

The main conclusion made in the paper is
that, in the case when closed time-like
lines exist, a sort of superselection (decoherence) may arise
in certain conditions,
i.e. the superposition principle may be partially violated. In
this case the conventional unitary description is not applicable
in the time-machine spacetime. Instead, description by a series
of partial evolution operators, obeying the generalized unitarity
condition, must be applied. The description of this type has
been earlier worked out in quantum theory of
continuous measurements.

It is well known now that the superposition principle of quantum
mechanics is restricted when a measurement is performed. One
may say that a sort of superselection exists in the situation of
measurement, forbidding to superpose the states corresponding to
different superselection sectors. In the case of quantum
measurements the superselection sectors correspond to different
alternative measurement outputs. The term
`decoherence' is often applied to describe this situation
\cite{Zeh73}-\cite{JoosZeh85}.

The superselection (decoherence) takes a specific form if a
continuous (prolonged in time) measurement is performed (see
\cite{Men79a}-\cite{Men-book3} and references therein). In the
procedure of path integration, those paths that are connected
with different alternative outputs $\al$ of the continuous
measurement, cannot be summed up. Instead of this, the amplitude
$A_{\al}$ of each of these alternatives must be calculated by
summation of the paths corresponding to the alternative $\al$.
After
this, the probability of each alternative $\al$ can be found as a
square modulus of the corresponding amplitude $A_{\al}$. The sum
of all these probabilities should be equal to unity.

This is valid not only in the situation when the measurement is
performed on purpose, but also when a measurement-like
interaction
of the system of interest with its environment takes place. The
latter means that information is recorded in the environment
about the state of the system or its evolution. This information
may be described with the help of alternatives. Each alternative
$\al$ is a class of states or a class of paths of the system.
Different classes should be considered as classical (decohering)
alternatives. Superposition of amplitudes corresponding to
different alternatives $\al$ is forbidden.

In practice the situation of the type of a measurement arises
each time when ``macroscopically distinct'' states (or ways of
evolution) of the system exist. What states must be considered
to be macroscopically distinct, depends on what environment the
system has.

This dependence on environment may be illustrated by the
well-known two-slit experiment. Propagation of the particle
through the first or the second slit suggests two alternatives.
Usually these are considered as quantum (`coherent',
`interfering') alternatives. One cannot know what of this
alternative is actually realized. In this case superposition of
paths describing propagating through different slits is possible,
leading finally to the well-known interference effect.

However this consideration is valid only when the slits are in
vacuum or in such a medium which cannot distinguish between two
alternatives. If the interaction with the medium (environment)
records information in this medium about what slit the particle
propagates through, the situation is quite different and the
preceding consideration is incorrect. It is possible of course to
consider the environment and its interaction with the particle
explicitly. However the influence of the environment onto the
particle may be taken into account implicitly if one consider two
alternatives (corresponding to the slits) to be classical
(`non-coherent', `incompatible'). Superposing the paths passing
through different slits is in this case forbidden. This is the
basic idea of quantum theory of measurements. More complicated
situations of continuous quantum measurements may be found in
\cite{Men-book3}.

Propagating in a spacetime containing CTC's, a particle may
travel backward in time. This means that it can pass through some
time interval (dischronal region) twice or more. Let the number
of times the particle returns to its past be $n$. Then we have
different ways of propagation characterized by the number $n$ (in
fact, nothing else than a winding number arising from
non-trivial topology of the time-machine spacetime).

The situations characterized by different numbers $n$ are quite
different from the physical point of view. For the given $n$ the
particle passes through the dischronal region (within the time
machine) $n+1$ times. An observer in the dischronal region
(within the time machine) will see $n+1$ particles existing
simultaneously. Different values of $n$ correspond to different
numbers of particles in the dischronal region.

However in non-relativistic quantum mechanics the number of
particles is conserved, and
a superselection corresponding to this number usually arises.
If it really arises,
the states corresponding to different numbers of
particles may not be superposed.

The superselection (decoherence) connected with the number of
particles is not absolute. It takes place for a massive particle
possessing a charge of some type (for example electric charge) or
spin (spinor charge). Even for such particles the decoherence
exists only ``in normal conditions''. The
latter means that the environment (consisting for example of
photons) distinguishes between the states with different numbers
of particles. Then any superposition of such states will decohere
very quickly so that the superpositions may be considered
to be forbidden, not existing.

This may be invalid in special conditions. Superpositions of
the states with different numbers of particles exist for example
in the superfluid (liquid helium at low temperature). In this
case the environment does not distinguish between different
numbers of particles, and no superselection connected with
the particle number exists.

In all considerations of the present paper we shall suppose that
the conditions in the dischronal region of the time machine are
``normal'' in this sense, i.e. they lead to superselection.
Then the number of particles (of the
considered type) existing simultaneously in this region presents
a classical alternative. If this number is equal to $n+1$, this
means that the particle $n$ times returned from the future to the
past. The situations corresponding to different $n$ should be
treated as non-coherent.

This presumption means that for the environment (medium)
within the time
machine alternative ways of propagation of the particle
corresponding to different $n$ are macroscopically
distinct. One
may say that a sort of measurement is performed in this case, and
the result of the measurement is characterized by the number $n$.
According to what has been said above, the paths characterized by
different winding numbers $n$ should be considered as
non-coherent (`decohering'). Their summation is forbidden.
Only the paths corresponding
to the same $n$  may be summed up forming corresponding
propagators and evolution operators $U_n$.\footnote{For evolution
into or from the dischronal region summing over the $n$th class
of paths is insufficient, and multiplication by a certain projector
is needed to obtain $U_n$, see Sect.~\ref{unitar-2}.} Propagators
and evolution operators with different $n$ should not be summed up.

Hence, in the case when the superselection of
different $n$ takes place, the ``general'' evolution operator $U
= \sum_n U_n$ (equal to the sum over all paths) makes no sense,
and the evolution of the particle in a time-machine spacetime
must be described in terms of the `partial evolution operators'
$U_n$. We shall develop this way of description including proof
of generalized unitarity $\sum_n {U_n}^{\dagger} U_n = {\bf 1}$
and generalized multiplicativity
$U_m(t'',t')\,U_n(t',t)=U_{m+n}(t'',t)$. The evolution in a
chronal region, between two chronal regions and within the
dischronal region (within the time machine) will be considered.
The resulting theory will be compared with the conventional
(coherent) description of evolution as given in \cite{Th94}.

The paper is organized in the following way. In
Sect.~\ref{sect-propagat} general information about propagators
and evolution operators in conventional theory is given, and
an analogue of this formalism is
presented (taken from quantum theory of measurement) for the case
when the superposition principle is restricted by superselection.
In Sect.~\ref{classes} the superselection sectors (coherency
sectors) are defined corresponding to the topological classes of
particle trajectories in the time-machine spacetime. On the basis
of this definition in Sects.~\ref{unitar-1},~\ref{unitar-2} the
generalized unitarity and in Sect.~\ref{sect-mult} the
generalized multiplicativity of evolution in the time-machine
spacetime is proven. In Sect.~\ref{discuss} the results obtained
are compared with the conventional unitary description of
evolution in the time-machine spacetime. Special kind of states
`trapped' within the time machine are described in
Sect.~\ref{trapped}. Sect.~\ref{conclus} contains short
concluding remarks.

\section{Propagators, Paths and Measurements}\label{sect-propagat}

In the conventional non-relativistic quantum mechanics
the propagator $K(t'',x''|t',x')$ is a probability amplitude for
the particle to transit from the point $x'$ at time moment $t'$
to the point $x''$ at time moment $t''$. The propagator may be
considered as a kernel of the evolution operator $U=U(t'',t')$:
\be
\psi_{t''} = U(t'',t')\,\psi_{t'}, \quad
\psi_{t''}(x'')
= \int K(t'',x''|t',x')\,\psi_{t'}(x')\, dx'.
\ee
An explicit expression for the propagator is presented by the
path integral,
\be\label{path-int}
K(t'',x''|t',x')
= \int d[x] \exp\left( \frac{i}{\hbar} S[x] \right),
\ee
where integration is performed over all paths
$[x] = \{ x(t)| t'\le t \le t'' \}$
connecting the points $(t',x')$ and $(t'',x'')$, and $S[x]$ is
the action functional calculated along the path $[x]$.

The evolution operator (propagator) satisfies the equation
\be\label{mult}
U(t'',t')\,U(t',t)=U(t'',t), \quad
\int K(t'',x''|t',x')\,K(t',x'|t,x)\, dx' = K(t'',x''|t,x)
\ee
(for $t \le t' \le t''$) which we shall call the property
of {\em multiplicativity}.
Besides, the evolution should conserve scalar products of the
states, thus the evolution operator should be {\em unitary}
that may be written in terms of propagators as follows:
$$
\int K^{*}(t'',x''|t',x')\,K(t'',x''|t,x)\, dx'' =
\delta (x',x).
$$

The formalism of propagators and evolution operators
must however be modified if a
measurement or an observation of the system is performed.
Such a modification is suggested by quantum theory of
measurements and particularly by quantum theory of continuous
measurements (see \cite{Men79a}-\cite{Men-book3} and
references therein).

The main feature of the resulting formalism is that a set of
classical alternatives $\al$ should be considered. These
alternatives arise as different measurement outputs. However,
for emerging the situation of this type, it is not necessary
that a measurement or an observation be arranged on purpose.
The only condition necessary for this is that some information
about quantum system be recorded in classical form in its
environment.
In the case of the time machine, the topological classes of
trajectories described in Sect.~\ref{classes} will play the
role of classical alternatives.

The alternatives $\al$ are classical in the sense that they are
incompatible. Correspondingly, quantum amplitudes corresponding
to different alternatives cannot be summed up: they are
non-coherent. Particularly, instead of a single propagator
$K(t'',x''|t',x')$ or a single evolution operator $U(t'',t')$ a
series of partial propagators  $K_{\al}(t'',x''|t',x')$ or
partial evolution operators  $U_{\al}(t'',t')$ are necessary for
describing evolution of the system. The partial propagators have
not to be summed up.

How the partial evolution operators may be used to describe the
evolution of the measured system? This should be made in
different ways in the case of selective situation (for example
selective measurement) and non-selective one. Selective
situation means that it is known what alternative $\al$ is
realized. The evolution
is described in this case by one partial propagator or evolution
operator $U_{\al}(t'',t')$. The evolution law, in terms of the
wave function (state vector) or the density matrix, is following:
\be\label{select-evolut}
\psi^{\al}_{t'} =
U_{\al}(t',t)\,\psi_{t},\quad
\rho^{\al}_{t'} =
U_{\al}(t',t)\,\rho_{t}\left( U_{\al}(t',t) \right)^{\dagger}.
\ee

The non-selective situation means that it is not known what
concrete alternative is realized. In this case one should sum up
over all possible alternatives. However, dealing with classical
alternatives, one must sum up probabilities, not amplitudes.
This means that summing must be performed in the second of the
formulas (\ref{select-evolut}) resulting in the following
evolution law:
\be\label{non-select-evolut}
\rho_{t'} = \sum_{\al}\rho^{\al}_{t'}
= \sum_{\al}U_{\al}(t',t)\,
\rho_{t}\left( U_{\al}(t',t) \right)^{\dagger}.
\ee

The final density matrix $\rho_{t'}$ has to be normalized ($\tr
\rho_{t'} = 1$)  for any normalized initial matrix $\rho_{t}$.
This takes place if and only if the following condition is
fulfilled:
\be\label{gen-unitar}
\sum_{\al}\left( U_{\al}(t',t) \right)^{\dagger} U_{\al}(t',t)
= {\bf 1}.
\ee
It may be called the {\em generalized unitarity condition}. This
condition means conservation of probability provided the
probability of the alternative $\al$ to belong to the set
$\cal A$ of alternatives is defined as follows (see
\cite{Men-book3}):
\be\label{prob}
{\rm{Prob}}(\al\in{\cal A}) =
\tr\sum_{\al\in{\cal A}}\rho^{\al}_{t'}
= \tr\sum_{\al\in{\cal A}}U_{\al}(t',t)\,
\rho_{t}\left( U_{\al}(t',t) \right)^{\dagger}.
\ee

If the alternatives (measurements) $\al_{t'}^{t''}$
corresponding
to different time intervals $[t',t'']$ may be considered,
their multiplication may be introduced,
$$
\beta^{t''}_{t'}\, \al^{t'}_{t} = \gamma^{t''}_{t}.
$$
Then the following {\em multiplicative law} should be valid
\cite{Men-book3} for the corresponding evolution operators
(propagators):
\be\label{mult-altern}
U(\beta^{t''}_{t'})\,U(\al^{t'}_{t})
=U(\gamma^{t''}_{t}), \quad t \le t' \le t''.
\ee

All these concepts, elaborated previously for continuous
quantum
measurements, may now be applied to the alternatives
corresponding to different topological numbers $n$ in the
time-machine spacetime.

In the paper \cite{Th94} the formalism of propagators in the
form of path integrals has been applied to describe the
evolution of a non-relativistic particle in the time-machine
spacetime. The propagator for such a particle was defined
as an integral over {\em all paths}.
It was shown that, in a simple model of the
time-machine spacetime, unitarity and multiplicativity for
the propagator are
violated if at least one of the time moments is in the
dischronal time interval.

We argued in Sect.~\ref{intro} that a sort of superselection
may arise in the time-machine spacetime for topologically
different paths. If this is the case,
then the ``general'' propagator
determined by integrating over all paths makes no sense.
Instead, one must use {\em partial propagators} just as in
theory of continuous quantum measurements. Each of
partial propagators should be a sum over
topologically equivalent paths.\footnote{This is valid for the
evolution from the past of the time machine to its future. For
evolution into or from the time machine the sum over
topologically equivalent paths should be multiplied by a
certain projector, see Sect.~\ref{unitar-2}.} Let us come
over to description of the corresponding classes of paths and
partial propagators.

\section{Topological Classes of Paths}\label{classes}

Let us consider the concrete model of a time-machine spacetime
and define the topological classes of trajectories of a
non-relativistic particle in such a spacetime. We shall discuss
hereafter the model of a time-machine spacetime that has been
used in the paper \cite{Th94}. It will be evident though
that at least some of the results have
more general validity.

The spacetime under consideration may be constructed (see
Fig.~\ref{fig-tm}) from a usual, `chronal' one by adding two
temporary wormholes. The original chronal spacetime will be
called the {\em background spacetime}. The wormholes added to the
background spacetime connect two space regions $S_1$ and $S_2$
belonging to the time slices $t_1$ and $t_2$ correspondingly,
with $t_1\le t_2$. One of the wormholes $W_1$ leads from the past
region $S_1$ to the future region $S_2$. This wormhole is entered
if the region $S_1$ is approached from its `past side' (but not
through the wormhole $W_2$). Another wormhole $W_2$ leads from
the future region $S_2$ to the past region $S_1$ if the region
$S_2$ is approached from its `past side' (but not through the
wormhole $W_1$). About the properties of walls of the wormholes
see the paper \cite{Th94}.
\begin{figure}
\vspace{5 cm}
\caption{\rm The time-machine spacetime is constructed from an
usual (chronal) background spacetime by attaching two wormholes
connecting two space regions in two different time slices. The
wormhole $W_1$ leads from the past to the future connecting the
`past
side' of the region $S_1$ with the `future side' of the region
$S_2$. The wormhole $W_2$ leads from the future to the past
connecting the `past side' of the region $S_2$ with the
`future side' of the region $S_1$.}
\label{fig-tm}\end{figure}

Consider all paths of a non-relativistic particle in this
spacetime and divide them in the topological classes. The
classes will then be connected with non-coherent sectors
(classical alternatives) in description of the particle
evolution in the spacetime.

The first class is formed by the paths going
through the `future-directed' wormhole $W_1$, entering $S_1$ and
going out from $S_2$. Let us denote this class by
${\cal P}_{00}$. The scheme of the path belonging to this class
is following:
$$
{\cal P}_{00}:\mbox{ initial point} \rightarrow S_1 \rightarrow
W_1 \rightarrow \mbox{final point}.
$$

One more class ${\cal P}_{0}$ includes the paths lying completely
in the background spacetime. These paths bypass both
entrances $S_1$ and $S_2$
to the wormholes. This is a class of topologically simple paths.
The scheme of the paths of this class is following:
$$
{\cal P}_{0}:\mbox{ initial point} \rightarrow \mbox{final point}.
$$

The class  ${\cal P}_{1}$ contains the paths that pass through
the wormhole $W_2$ once. The path of this type enters the
mouth $S_2$ of the wormhole at time $t_2$ and go out of the mouth
$S_1$ at time $t_1$. The scheme of these paths is following:
$$
{\cal P}_{1}:\mbox{ initial point} \rightarrow S_2 \rightarrow
W_2 \rightarrow \mbox{final point}.
$$

The path of the class ${\cal P}_{2}$ proceeds through the
wormhole
$W_2$ twice, going into $S_2$ at time $t_2$, going out of
$S_1$ at $t_1$, proceeding through the background spacetime
again to time $t_2$, once more entering $S_2$, again going
out of $S_1$, and ultimately approaching the final point
through the background
spacetime. This corresponds to the following scheme:
$$
{\cal P}_{2}:\mbox{ initial point} \rightarrow S_2 \rightarrow
W_2 \rightarrow S_2 \rightarrow W_2 \rightarrow
\mbox{final point}.
$$

Generally, the paths of the class  ${\cal P}_{n}$  with
$n\ge 0$ pass through the wormhole $W_2$ precisely $n$
times, according to the scheme
$$
{\cal P}_{n}:\mbox{ initial point}\; \big( \rightarrow S_2
\rightarrow W_2 \big)^n\rightarrow \mbox{final point}.
$$

It was argued in Introduction that the types of evolution
corresponding to different numbers of returns
of the particle to its past are usually (in ``normal
conditions'')
macroscopically distinct and therefore cannot be coherently
superposed. These types of evolution correspond to the paths
belonging to different topological classes. If the decoherence
takes
place, the propagator defined by integrating over all paths (as
in Eq.~(\ref{path-int})) makes no sense. It seems reasonable to
consider instead the propagators obtained by integrating over
the topological classes:
\be\label{path-int-class}
K_n(t'',x''|t',x') = \int_{{\cal P}_{n}} d[x] \exp\left(
\frac{i}{\hbar} S[x] \right).
\ee
The operators corresponding to these two-point functions will be
denoted by $K_n$:
$$
\psi_{t''}(x'') = \big( K_n(t'',t')\,\psi_{t'} \big)\, (x'')
= \int K_n(t'',x''|t',x')\,\psi_{t'}(x')\, dx'.
$$

If both the initial and final time moments $t',t''$ are in the
same chronal region (before or after the time machine), then only
one topologically trivial class $n=0$ exists, and only one
operator $K_0$ is defined by Eq.~(\ref{path-int-class}).
It is evident that,
according to general theory (Sect.~\ref{sect-propagat}), this
operator $K_0=U_0=U$ is the evolution operator of the particle in
the chronal region. This operator is unitary.

If the time moments $t',t''$ are in different chronal regions or
some of them (or both) are in the dischronal region, then there
are operators $K_n$ corresponding to arbitrary
$n=00,0,1,2,\dots$. They seem to be good candidates for
describing evolution in the $n$th superselection sector ($n$th
classical alternative). We shall see that this is valid in the
case when the time $t'$ is before the time machine emergence,
$t'<t_1$ and the time $t''$ is after it disappearing, $t''>t_2$.
In this case the operators $K_n=U_n$ play the role of `partial
evolution operators' satisfying the generalized unitarity
condition, $\sum_n {U_n}^{\dagger}U_n=1$. However when the
time moment $t''$ is in the dischronal region
(within the time machine), then the partial evolution operators
$U_n$ will be shown to differ from the operators $K_n$ by
certain projectors (see Sect.~\ref{unitar-2}).

This is only natural because of the following.

The coherency sectors for evolution of a particle in the
time-ma\-chine spacetime may be defined as a number of times the
particle returns to its past. The set of such superselection
sectors is in one-to-one correspondence with the topological
classes ${\cal P}_n$ of paths between the time $t'<t_1$ and the
time $t''>t_2$. Thus the above-defined operators $K_n$ must
coincide in the case $t'<t_1<t_2<t''$ with what we call `partial
evolution operators' $U_n$. However, if the final time $t''$ of
the evolution is within the time machine,
$t_1<t''<t_2$, the class ${\cal P}_n$ of paths leading to this
time does not correspond to the $n$th coherency sector.

Indeed,
after the time moment $t''$ the particle can either escape the
time machine or enter the wormhole $W_2$ and return to its past
one or more times. Therefore, the operator $K_n$ (defined by
summing up over the class ${\cal P}_n$) does not in fact coincide
with the partial evolution operator for the $n$th alternative. If
however we provide, multiplying $K_n$ by the corresponding
projector, that the particle escape the time machine after
$t''$, then we have the correct partial evolution operator $U_n$.

\section{Generalized Unitarity I}\label{unitar-1}

We considered in Sect.~\ref{sect-propagat} the general situation
when non-coherent sectors exist in the evolution
of a quantum system. In Sect.~\ref{classes} coherency sectors
connected with topological classes of paths were defined in
the time-machine spacetime.
In the present section and the next one we shall formulate and
prove the generalized unitarity condition for this case.

As has already been argued, the classical alternatives may be
identified (at least for the case $t'<t_1<t_2<t''$) with the
classes of paths ${\cal P}_n$ introduced in Sect.~\ref{classes}.
Operators $K_n$ were defined in Sect.~\ref{classes} by summation
over classes ${\cal P}_n$. It is natural to try and identify
these operators with the partial evolution operators $U_n$
corresponding to the alternatives. This turns out to be correct
for the
most important case $t'<t_1<t_2<t''$ (evolution from the past of
the time machine to its future). This may be accepted also in the
cases when both $t',t''$ are in the same chronal region. A more
complicated situation arises if $t''$ is within the time machine
(in the dischronal region). The corrections for this case will be
introduced in the next section.

Consider first the trivial situations of the evolution within
one of two chronal regions. In the case $t'<t''<t_1$ (as well
as for $t_2<t'<t''$) there is only one (trivial) class of paths
with $n=0$ and only one operator $K_0$. This operator coincides
in fact with the evolution operator in the background spacetime.
Evolution in this case does not differ from usual coherent
(unitary) evolution and is described by a single evolution
operator $U=U_0=K_0$ (we omit hereafter the explicit
specification of the initial and final time moments).
The generalized unitarity condition
coincides in this case with the conventional unitarity,
$U^{\dagger}U={\bf 1}$.

If the evolution not restricted by a single
chronal region is considered, all topological classes
$n=00,0,1,2,\dots$
(and corresponding coherency sectors) exist.
According to the general formula (\ref{gen-unitar}), the
generalized unitarity condition must then have the form
\be\label{tm-unitar}
{U_{00}}^{\dagger}U_{00} + \sum_{n=0}^{\infty}
{U_n}^{\dagger}U_n = {\bf 1}
\ee
where we omitted specification of the time interval. We shall
see that this condition is actually fulfilled for the evolution
operators $K_n$ corresponding to the propagators
(\ref{path-int-class}) provided the final time of the evolution,
$t''>t_2$. This is why the operators $K_n$ may actually be
identified with the partial evolution operators $U_n$ in this
case.

The generalized unitarity condition (\ref{tm-unitar}), as
usual for quantum mechanics, expresses conservation of
probability.
If an initial state is described by the density matrix $\rho$,
then the probability of realization of $n$th alternative is,
according to quantum theory of measurements (see
Eq~(\ref{prob})),
\be\label{nth-prob}
P_n = \tr \rho_n = \tr \left( U_n \rho {U_n}^{\dagger} \right).
\ee
The condition (\ref{tm-unitar}) provides then that
$$
P_{00} + \sum_{n=0}^{\infty} P_n = 1
$$
for an arbitrary initial state $\rho$. The same probability
interpretation of the generalized unitarity is valid for all
other choices of $t'$, $t''$ that will be considered below.

To demonstrate that the condition (\ref{tm-unitar}) is valid,
we shall express the operators $K_n$ in a more explicit form,
through the evolution operators in the background chronal
spacetime (we shall denote them by $V_i$) and the evolution
operators in the wormholes (denote these operators $W_i$).
Besides this, two projectors
in the background spacetime ($P_i$) will be used describing
entering the mouths of the wormholes. Expressions for the
operators $K_n$ may be readily constructed with the help of the
description of classes ${\cal P}_n$ (see Sect.~\ref{classes}).

Consider first the case $t'<t_1<t_2<t''$, i.e. the
evolution starting before the time machine emergence and
finishing after it disappearance.

The paths of the class ${\cal P}_{00}$ go from the starting point
to the point at time moment $t_1$, then enter the wormhole $W_1$,
proceed through it to the moment $t_2$ and then go from $t_2$ to
the final point. Summation over all paths of this type gives us
the evolution operator $K_{00}$ in the form of the product $V_2
W_1P_1 V_1$ where $V_1$ and $V_2$ are the evolution operators
outside the time machine (first of them between starting time
$t'$ and time $t_1$, and the second one between $t_2$ and the
final time $t''$), and $W_1$ is the evolution operator along the
wormhole $W_1$. The projector $P_1$ onto the region $S_1$
provides entering the particle into the wormhole.

Analogously the expressions for all operators $K_n$ can be found
in the case $t'<t_1<t_2<t''$. We shall see below that these
operators satisfy the generalized unitarity condition and may be
identified with the partial evolution operators. This is why we
shall denote them by $U_n$:
\be\label{out-time}
U_{00} = K_{00} = V_2 W_1P_1 V_1, \quad
U_{n} = K_{n} = V_2 (1-P_2) (V_{21} W_2 P_2)^n V_{21}(1-P_1)
V_1, \quad n\ge 0.
\ee
Here $V_{21}$ is the evolution operator in the background
spacetime between time moments $t_1$ and $t_2$. The projector
$P_2$ provides entering the region $S_2$ and therefore
the wormhole $W_2$ while the complementary projector $1-P_2$
provides bypassing the wormhole $W_2$. Analogously the projector
$1-P_1$ provides bypassing $W_1$.\footnote{We denoted here the
unity operator by $1$ instead of ${\bf 1}$.}

Now we can easily prove the generalized unitarity
(\ref{tm-unitar}) for the operators $U_n=K_n$ defined by
Eq.~(\ref{out-time}). In the proof we shall use
unitarity of all $V_i$ and $W_i$ as well as the properties of
projectors, $P_i^2=P_i$ and ${P_i}^{\dagger}=P_i$. Let us
present the expression ${U_n}^{\dagger}\,U_n$ in the form
\ba\label{Un+U}
{U_n}^{\dagger}\,U_{n}
&=& A^{\dagger} \left( B^{\dagger}\right )^n
{V_{21}}^{\dagger}(1-P_2) V_{21}  B^n A \nonumber\\
&=& A^{\dagger} \left( B^{\dagger}\right )^{n-1} C  B^{n-1} A
- A^{\dagger} \left( B^{\dagger}\right )^n C  B^n A
\ea
where
$$
A=  (1-P_1) V_1, \quad
B=  W_2 P_2 V_{21}, \quad
C= {V_{21}}^{\dagger}P_2 V_{21} .
$$
Summing (\ref{Un+U}) in $n$, we shall obtain
\be
\sum_{n=1}^{\infty} {U_n}^{\dagger}\,U_{n}
= A^{\dagger}  C   A
= {V_1}^{\dagger} (1-P_1)
{V_{21}}^{\dagger}P_2
V_{21}  (1-P_1) V_1.
\ee
Adding two other terms ${U_0}^{\dagger}U_0$ and
${U_{00}}^{\dagger}U_{00}$, we shall prove finally the
generalized unitarity condition (\ref{tm-unitar}).

\begin{remark}\label{fine-tune} {\rm
In the proof given above (and in the analogous proofs hereafter)
it is supposed implicitly that the terms of the sum
(\ref{tm-unitar}) tend to zero when $n$ tends to infinity. This
is not valid in the subspace of some `finely tuned' states, and
the formula (\ref{tm-unitar}) is not applicable in this subspace.
More precisely, this formula, as an equality of two operators, is
valid in the space of states $|\psi\rangle$ for which
$$
\lim_{n\to \infty}\langle\psi |
A^{\dagger} \left( B^{\dagger}\right )^n C  B^n A
|\psi\rangle = 0
$$
(in the notations introduced above). From the physical point of
view, this means that the probability for the particle (in the
considered state) to return $n$ times into its past infinitely
decreases when $n$ tends to infinity. This is right for `almost
all' states. We shall see in Sect.~\ref{trapped} however that
this is not valid for some finely tuned states trapped within the
time machine.}
\end{remark}

Consider now the case when the initial time moment $t'$ is within
the interval $[t_1,t_2]$, but the final time $t''$ is after
$t_2$. In this case the paths cannot belong to the class ${\cal
P}_{00}$, thus the generalized unitarity takes the form
\be\label{tm-unitar-in}
\sum_{n=0}^{\infty} {U_n}^{\dagger}U_n = {\bf 1}.
\ee
Let us prove it.

In the case $t_1<t'<t_2<t''$ we have for the operators $K_n$
(that again will prove to coincide with the partial evolution
operators $U_n$) the following formulas:
\be\label{in-past-time}
U_{n} = K_{n}
 = V_2 (1-P_2) (V_{21} W_2 P_2)^{n} V_{20},
\quad n\ge 0,
\ee
where $V_{20}$ denotes the propagator in the background (chronal)
spacetime from the initial time $t'$ (between $t_1$ and
$t_2$) to the time $t_2$.
Summing up the expression for ${U_n}^{\dagger}\,U_{n}$ in the
same way as above gives the
generalized unitarity in the form (\ref{tm-unitar-in}) provided
we accept the definition (\ref{in-past-time}).

Let now the final
time $t''$ of the evolution be between $t_1$ and $t_2$. We shall
see that the (generalized) unitarity does not take place for the
operators $K_n$ in this case, so that these operators cannot be
identified with partial evolution operators. Right
form of the partial evolution operators will be given
in Sect.~\ref{unitar-2}.

Constructing the operators $K_n$ in the same way as above
(i.e. as integrals over the topological classes of paths),
we are led to the following formulas for
the case $t'<t_1<t''<t_2$ (when the initial time is earlier
than the time machine emergence):
\be\label{in-fut-time}
K_{n} = V_{01} (W_2 P_2 V_{21})^n (1-P_1) V_1, \quad n\ge 0
\ee
($V_{01}$ describes evolution between $t_1$ and the final time
$t''$). If both initial and final times are within the time
machine ($t_1<t'<t''<t_2$), we have
\be\label{both-in}
K_{0} = V_{0'0}, \quad
K_{n} = V_{0'1} W_2 P_2 (V_{21}W_2 P_2 )^{n-1} V_{20},
\quad n\ge 1.
\ee
In both cases (\ref{in-fut-time}), (\ref{both-in}) the
generalized unitarity does not take place.

One can readily see that the algebraic operations used earlier
to prove the generalized unitarity, are now impossible because
there is no projector of the form $1-P_2$ in the expressions
(\ref{in-fut-time}), (\ref{both-in}). This gives a hint that
the operator $U_n$ must differ from $K_n$ by the factor of the
form $1-Q_2$ where $Q_2$ is a projector. We shall see in
Sect.~\ref{unitar-2} that this may be justified physically.

\section{Generalized Unitarity II}\label{unitar-2}

It has been shown in the preceding section that the operators
$K_n$  defined
with the help of the classes of paths ${\cal P}_n$ and expressed
by the formulas (\ref{in-fut-time}), (\ref{both-in}) do not
satisfy the generalized unitarity condition if the evolution is
considered to the time moment $t''$ between $t_1$ and
$t_2$.The physical reason of this is that the classes of paths do
not correspond in this case to the coherency sectors, i.e. to
macroscopically distinguishable (distinct) physical situations.
Therefore, the operators $K_n$ are not in this case partial
evolution operators.

Indeed, the situations are macroscopically distinct
if they differ by the number of particles in the time
interval $[t_1,t_2]$. If the path belongs to the class ${\cal
P}_n$ and ends after $t_2$, it describes a quite definite number
of particles, $n+1$ (because returning to the past is impossible
after time $t_2$). However, if the path ends in the point $t''<
t_2$, then the number of particles in the interval $[t_1,t_2]$ is
indefinite, since it is not known whether the particle will
escape the time machine after $t''$ or enter the back wormhole
$W_2$, return to its past and propagate through the time machine
once more. To fix the sector of coherency (classical
alternative), we have to introduce projectors $1-Q_2$ and $Q_2$
corresponding to the two possibilities described above: escaping
the time machine and entering it once more.

It is evident how this may be done. If the operator $V_{20}$
describes the evolution between the time moments $t''$ (the final
point of evolution) and $t_2$, then the operator
\be\label{Q_2}
Q_2 = V_{20}^{-1} P_2 V_{20} = {V_{20}}^{\dagger} P_2 V_{20}
\ee
projects on those states at time $t''$ which after the evolution
to the time $t_2$ will enter the region $S_2$ and return to the
past. Accordingly, the operator $1-Q_2$ distinguishes the states
corresponding to the particle escaping from the time machine.

Having the projector $1-Q_2$ and using Eq.~(\ref{in-fut-time}),
one can find the partial evolution operators for the case
$t'<t_1<t''<t_2$:
\be\label{in-fut-time-Q_2}
U_{n} = (1-Q_2)K_n = (1-Q_2)V_{01} (W_2 P_2 V_{21})^n (1-P_1)
V_1, \quad n\ge 0.
\ee
Let us remark that, in a sense, these operators, not
(\ref{in-fut-time}), correspond to the topological classes of
paths ${\cal P}_n$. Indeed, they actually fix the number of times
the particle returns to its past, not only before time $t''$, but
also after this. The operators (\ref{in-fut-time-Q_2}) provide
that the paths from the only one class ${\cal P}_n$ contribute
the evolution of the particle after $t''$ as well as before this
time.

Expressing $Q_2$ in (\ref{in-fut-time-Q_2})
through $P_2$ (due to (\ref{Q_2})) and using the equation
$V_{21}=V_{20}V_{01}$ (multiplicativity for $V_{21}=V(t_2,t_1)$),
we have finally the following formula for the partial evolution
operators in the case $t'<t_1<t''<t_2$:
\be\label{{in-fut-time-P_2}}
U_{n} = {V_{20}}^{\dag} (1-P_2)(V_{21}W_2 P_2)^n V_{21}(1-P_1)
V_1, \quad n\ge 0.
\ee

Let us try to prove the generalized unitarity for these
partial evolution operators. One may for example
substitute $P_2$ in the formula (\ref{in-fut-time-Q_2}) by
its expression through $Q_2$ to get
\be\label{in-fut-Q_2-mod}
U_{n} = (1-Q_2)V_{01}
(W_2 V_{20} Q_2 V_{01})^n
(1-P_1) V_1, \quad n\ge 0.
\ee
Then, acting just as in Sect.~\ref{unitar-1}, we have
\be\label{in-fut-incompl}
\sum_{n=0}^{\infty} {U_n}^{\dagger} U_n =
{V_1}^{\dagger}(1-P_1)V_1.
\ee

The generalized unitarity is not yet fulfilled.
The reason is that one more partial evolution operator must be
considered, describing one more channel of evolution. Indeed,
if the initial time is before $t_1$, then the particle have
the possibility to enter the forward-directed wormhole $W_1$.
The probability of
this alternative is to be taken into account.

To take it into account, we must interpret
the concept of a `final time moment' in another way.
So far we supposed that the
`final time' is nothing else than a space-like surface
($t''=\mbox{const}$) {\em in the background spacetime}
(the spacetime without wormholes). However when the final
time is within the time machine (i.e. later than $t_1$ but
earlier than $t_2$) this `time' may be thought of as a sum
of the space-like surface in the background spacetime {\em plus}
a space-like {\em slice of the wormhole} $W_1$.

With this wider concept of the final moment, the particle may
arrive at this moment
not only through the background spacetime but also through the
wormhole $W_1$. The first possibility has been taken into account
by the operators (\ref{in-fut-Q_2-mod}). The latter possibility
is described by the operator
\be\label{W1}
U_{00} = W_{01} P_1 V_1.
\ee
Adding the term
${U_{00}}^{\dagger}U_{00}$ to the sum (\ref{in-fut-incompl})
gives, in the case $t'<t_1<t''<t_2$, the generalized
unitarity condition in the form of Eq.~(\ref{tm-unitar}).

At last, consider the case $t_1<t'<t''<t_2$ when both initial
$t'$ and final $t''$ time moments are within the interval
$[t_1,t_2]$. Using the partial evolution operators obtained from
(\ref{both-in}) by inclusion the projectors
$1-Q_2={V_{20}}^{\dag}(1-P_2)V_{20}$, we have
\be \label{both-in-Q2}
U_{n} = {V_{20'}}^{\dag}(1-P_2) (V_{21}W_2 P_2 )^{n}
V_{20}, \quad n\ge 0,
\ee
we shall easily prove the generalized unitarity in this case
too:
\be\label{both-in-unitar}
\sum_{n=0}^{\infty} {U_n}^{\dagger} U_n = {\bf 1}.
\ee

One more remark may be made in connection with the argument
preceding Eq.~(\ref{W1}). If the concept of
time moment within the time machine is changed (as was
discussed in the above-mentioned argument), the same wider
concept must be
applied not only to `final', but also to the
`initial time moment'. Then, besides the operators
(\ref{both-in-Q2}), one more operator should be added,
\be\label{in-wormh}
U_{00} = W_{0'0}
\ee
describing evolution in the wormhole $W_1$ between two slices
of this wormhole.

It seems at the first glance that this additional operator
violates the (generalized) unitarity. One can see however that
this is not the case. Indeed, by the unity operator ${\bf 1}$
in Eq.~(\ref{both-in-unitar}), just as in all preceding formulas,
we meant the unity operator in the space of states
localized on the space-like surface of the background spacetime.
If we express this in the notation explicitly,
Eq.~(\ref{both-in-unitar}) should read
\be\label{unitar-out}
\sum_{n=0}^{\infty} {U_n}^{\dagger}U_n
= {\bf 1}_{\mbox{background}}.
\ee
The operator (\ref{in-wormh}) satisfy the relation
\be\label{unitar-in}
 {U_{00}}^{\dagger}U_{00} = {\bf 1}_{\mbox{wormhole}}
\ee
where the r.h.s. is the unity operator for the
states
localized on the space-like slice of the wormhole $W_1$. Summing
up both preceding formulas, one has the generalized unitarity in
the form
\be\label{unitar-compl}
{U_{00}}^{\dagger}U_{00} + \sum_{n=0}^{\infty} {U_n}^{\dagger}U_n
= {\bf 1}_{\mbox{complete}}.
\ee

Quite analogously, the operator of the form
\be\label{init-in-add}
U_{00}=V_2 W_{20}
\ee
may be added to the series of operators (\ref{in-past-time}).
The remarks analogous to those of the preceding paragraph
may be made in this case. The generalized unitarity takes then
the form (\ref{unitar-compl}) (instead of (\ref{tm-unitar-in}))
for the case $t_1<t'<t_2<t''$ too.

Thus, the generalized unitarity takes place for an arbitrary time
interval.  Therefore, the concept of probability may be used
correctly even within the time machine.

Let us make one more remark. We saw that an additional projecting
factor $1-Q_2$ should be including in the partial evolution
operators $U_n$ corresponding to the final time within the time
machine, $t_1<t''<t_2$. The same logic seems to require inclusion
of an analogous factor $1-Q_1$ in the expressions for $U_n$ in
the case when the initial time is within the time machine,
$t_1<t'<t_2$. This actually may be done. Then we guarantee that
the operators $U_n$ describe propagation of only those states
that have resulted from the evolution starting earlier than
$t_1$.
This however is not necessary because the operators $U_n$ may be
quite reasonably applied to arbitrary states prepared at time
$t'$ within the time machine. This is why the projector $1-Q_1$
turned out to be not necessary for providing the generalized
unitarity.

Preparation of the state within the time machine could be
restricted by the condition of self-consistency if interaction of
the particle with its duplicates were taken into account. We
however consider non-interacting particles (i.e. we suppose that
the interaction is negligible). Therefore, the state of {\em one
particle} at time $t'$ within the time machine may be prepared
arbitrarily.\footnote{Notice that the complete description of the
states of all particles at this moment (including the original
particle and its duplicates) is not arbitrary. Indeed, if we are
preparing, at time $t'$ between moments $t_1$ and $t_2$, the
state that will $n$ times return to its past, then it is already
known at $t'$ that $n$ duplicates of the particle exist coming
from the region $S_1$ at time $t_1$.} Further evolution of this
arbitrary state will be described by one of the partial evolution
operators $U_n$. Any number $n$ may be realized in this
evolution, the probability of each of them being determined by the
formula (\ref{nth-prob}).

\section{Multiplicativity for the Time Machine}
\label{sect-mult}

Let us address now the question of multiplicativity for the
evolution in the time-machine spacetime. Multiplicativity of
the propagators or evolution operators (\ref{mult}) means that
the evolution during some time interval may be considered in two
stages as evolution during two subintervals. With the decoherence
caused by measurement, the multiplicativity is described by the
formula (\ref{mult-altern}). We should now try and interpret this
formula for the evolution in the time-machine spacetime with the
specific decoherence (superselection) arising in this case.

The winding number $n$ introduced in Sect.~\ref{classes} through
the classes of paths ${\cal P}_n$ hints how multiplicativity
might be defined and proved in the case of the time machine.
Indeed, this number is equal to the number of times the particle
returns to its past travelling through the wormhole $W_2$. If we
have two alternatives characterized by the numbers $n_1$ and
$n_2$ correspondingly, then the product of these alternatives
corresponds to
the number $n=n_1+n_2$. This means simply that if the particle
returns $n_1$ times and then once more returns $n_2$ times to its
past, then ultimately it returns $n_1+n_2$ times. It is evident
that the (generalized) multiplicativity (\ref{mult-altern}) might
have, in the case of a time machine, the following form that can
be readily proved for the operators $K_n$:
\be\label{mult-tm}
K_{00}(t'',t')\,K_{00}(t',t) = K_{00}(t'',t), \quad
K_{m}(t'',t')\,K_{n}(t',t) = K_{m+n}(t'',t),
\quad n\ge 0.
\ee

The relation (\ref{mult-tm}) may be readily proved for the
operators $K_n$ with the help of the path-integral representation
(\ref{path-int-class}). The proof is based on the fact that each
path $p\in {\cal P}_{m+n}$ may be presented as a product of
the paths $p_1\in {\cal P}_{m}$ and $p_2\in {\cal P}_{n}$,
and vice
versa, the product of any paths $p_1\in {\cal P}_{m}$, $p_2\in
{\cal P}_{n}$ gives a path belonging to ${\cal P}_{m+n}$. By
`product' we mean passing through one of the paths and then
through another one.The product of two paths is defined only if
the first path ends in the point where the second one starts (see
\cite{Men-book3}, Chapter~10 for details of this algebra of
paths).

Instead of this, the relation (\ref{mult-tm}) for operators
$K_n$
may easily be proved by using explicit expressions for these
operators derived in Sect.~\ref{unitar-1}. The latter way of
consideration may be applied also to the partial
evolution operators $U_n$ which, as we know, do not always
coincide with the operators $K_n$. We shall see that for
partial evolution operators $U_n$
the second of the relations (\ref{mult-tm}) is valid only
when at least one of the numbers $m$, $n$ is zero.

It is evident that the generalized multiplicativity
(\ref{mult-tm}) is valid in the trivial cases $t'<t''<t_1$ and
$t_2<t'<t''$, when it reduces to the conventional
multiplicativity because both $m=n=0$:
$$
U_0(t'',t')\,U_0(t',t) = U_0(t'',t).
$$
Trivial is also the case when one of the intervals $[t,t']$ or
$[t',t'']$ lies in one of the chronal regions. Then the
multiplicativity takes one the following forms that
can be easily proven:
$$
U_n(t'',t) = U_0(t'',t')\,U_n(t',t)
\quad \mbox{or} \quad
U_n(t'',t) = U_n(t'',t')\,U_0(t',t).
$$

Consider now non-trivial situations.

Summing up the results of
Sects.~\ref{unitar-1},~\ref{unitar-2}, we have
the following formulas for the partial evolution operators
$U_{00}(t'',t')$, $U_n(t'',t')$ (where $n\ge 0$):
\ba\label{evol-all}
t'<t_1<t_2<t'': && \nonumber\\
U_{00} &=& V_2 W_1P_1 V_1, \quad
U_{n} = V_2 (1-P_2) (V_{21} W_2 P_2 )^n V_{21} (1-P_1) V_1.
\nonumber\\
t_1<t'<t_2<t'': &&\nonumber\\
U_{00} &=&V_2 W_{20}, \quad
U_{n} = V_2 (1-P_2) (V_{21} W_2 P_2)^{n} V_{20}.
\nonumber\\
t'<t_1<t''<t_2: &&\nonumber\\
U_{00} &=& W_{01} P_1 V_1,\quad
U_{n} = {V_{20}}^{\dagger}(1-P_2)(V_{21} W_2 P_2)^n
V_{21}(1-P_1) V_1. \nonumber\\
t_1<t'<t''<t_2: &&\nonumber\\
U_{00}  &=& W_{0'0}, \quad
U_{n} = {V_{20'}}^{\dagger}(1-P_2) (V_{21} W_2 P_2 )^{n}
V_{20}. \nonumber
\ea

One can verify straightforwardly that, of all
multiplicativity relations (\ref{mult-tm}), the following
are valid also for the partial evolution operators:
\be\label{mult-gen}
U_{00}(t'',t')\,U_{00}(t',t) = U_{00}(t'',t), \quad
U_{0}(t'',t')\,U_{n}(t',t) = U_{n}(t'',t),
\quad n\ge 0.
\ee
If $m\ne 0$, the second of the relations (\ref{mult-tm})
is not valid for $U_{n}$. Moreover, if we defined the
partial evolution operators with the time argument within the
time machine in such a way as it was supposed in the remark at
the end of Sect.~\ref{unitar-2},
then even the products of the form $U_0\,U_n$
with $n\neq 0$ would also give zero.

We see therefore that a topologically non-trivial evolution
(described by the evolution operator $U_n$ with $n\ge 1$) cannot
be presented as the product of two topologically non-trivial
evolutions. Topologically non-trivial evolution is `integral in
time', it cannot be followed step by step: first $m\neq 0$
returns to the past, then again $n\neq 0$ returns that finally
gives $m+n$ returns.

This is not astonishing. If we describe the evolution within
the time machine, not achieving an exit from it, then this part of
evolution cannot be characterized by a definite number $n$ of
returns of the particle to the past (this number depends on the
subsequent
stage of evolution). In a sense, evolution within the time
machine is not local in time, influence of the future cannot be
excluded.

We succeeded in constructing operators describing evolution to
the time moment $t''$ within the time machine, for example
$U_n(\mbox{in,past})$. However a special projector in this
operator provides escaping from the time machine after time
$t''$. This operator guarantees that the future stage of evolution
(after $t''$) will not influence the past evolution (until $t''$).
Thus, non-locality of evolution within the time machine is
evident even from the structure of partial evolution operators,
not only from the form of multiplicativity for them.

On the other side, the operators $K_n$ (but not $U_n$) possess the
property of multiplicativity (\ref{mult-tm}) for arbitrary time
arguments.

One may object that expressing multiplicativity in terms of the
operators $K_n$, in fact non-physical within the time machine,
makes no sense (if the superselection in $n$ takes place).
In fact, each of these operators describes
propagation of non-physical states (forbidden superpositions).
However, the product of such operators,
$$
K_{n_N}(t'',\tau_{N-1}) \dots K_{n_2}(\tau_2,\tau_1)
K_{n_1}(\tau_1,t')=K_{n_1+n_2+\dots +n_{N}}(t'',t'),
$$
with $t'<t_1<t_2<t''$, is a correct partial evolution operator.
No non-physical state is propagated due to this operator. Thus,
even though each of the operators $K_n$ describes propagation of
non-physical states within the time machine, the result of the
propagation from the past (in respect to the time machine) to
the future of the time machine is presented by these
operators correctly.

One may say that the operators $K_n$ give multiplicative
description of evolution even within the time machine, but at
the price of introducing non-physical states in
intermediate stages.

One more interpretation of the results obtained is
complementarity between (generalized) unitarity and
multiplicativity. One may describe the evolution within
the time machine by the operators $U_n$ (unitary in the
generalized sense but satisfying only trivial multiplicativity
relations) or by $K_n$ (satisfying the generalized
multiplicativity but not unitarity).

\section{Comparison with the Unitary Theory}\label{discuss}

We showed in Sects.~\ref{unitar-1},~\ref{unitar-2} that the
description of the particle evolution in the time-machine
spacetime with the help of the partial propagators (partial
evolution operators) is correct from the point of view of
probabilities. This statement means that the partial evolution
operators $U_n$ satisfy the condition of generalized unitarity
(\ref{tm-unitar}).

In the conventional theory, when no decoherence (superselection)
is supposed, the evolution is described by a single evolution
operator $U$ which must be unitary, $U^{\dagger}U=1$. In a number
of papers evolution of particles in a time-machine spacetime
was described in this way (see \cite{Th94} and references
therein). Let us compare two types of theories. This is not
difficult because we used here the same model of spacetime as
in \cite{Th94}.

It was shown in \cite{Th94} that a single evolution operator
$U=U(t'',t')$ is unitary in the case $t'<t_1<t_2<t''$ when the
evolution from the past of the time machine to its future is
considered. At the first glance, this contradicts to our
conclusion about generalized unitarity of the partial evolution
operators $U_n$ in this case.

Indeed, the evolution operator $U$ was defined in \cite{Th94} by
summation over all paths. In our terms, this means that this
operator is equal to
\be\label{sum-U_n}
U = U_{00} + \sum_{n=0}^{\infty} U_n
\ee
where $U_n$ are defined by Eq.~(\ref{out-time}).
Unitarity for this operator means that
\be\label{tm-unitar-Thorne}
{U_{00}}^{\dagger}U_{00}
+ \sum_{n=0}^{\infty} {U_{00}}^{\dagger}U_n
+ \sum_{n=0}^{\infty} {U_n}^{\dagger}U_{00}
+ \sum_{m,n=0}^{\infty} {U_m}^{\dagger}U_n = {\bf 1}.
\ee
The formulas (\ref{tm-unitar}) and (\ref{tm-unitar-Thorne})
differ by the non-diagonal terms (they are absent in the former
case). Both formulas may be valid only if a sum of non-diagonal
terms is zero. It turns out however that this actually takes
place in the considered case ($t'<t_1<t_2<t''$), so that both
forms of unitarity turn out to be valid. Let us show this with
the evolution operators (\ref{out-time}).

Consider an off-diagonal term in (\ref{tm-unitar-Thorne}),
${U_{n+k}}^{\dagger}U_n$, $k\ge 1$. Using  Eq.~(\ref{out-time}),
one may present this term in the form
\be\label{1}
{U_{n+k}}^{\dagger}U_n
=
A^{\dagger} (B^{\dagger})^{n+k-1} C B^{n-1} A
- A^{\dagger} (B^{\dagger})^{n+k} C B^{n} A
\ee
where it is denoted
$$
A=(1-P_1) V_1, \quad B=W_2 P_2 V_{21}, \quad
C={V_{21}}^{\dagger} P_2 V_{21}.
$$

The formula (\ref{1}) is valid for all $n\ge 1$. Summing up
this expression in $n$ from 1 to $\infty$ and adding the
terms ${U_{k-1}}^{\dagger}U_{00}$ and
${U_{k}}^{\dagger}U_{0}$, we have
\ban
\lefteqn{
{U_{k-1}}^{\dagger}U_{00} +
\sum_{n=0}^{\infty}{U_{n+k}}^{\dagger}U_n}\\
&=&
{V_1}^{\dagger} (1-P_1) ({V_{21}}^{\dagger} P_2
{W_2}^{\dagger})^{k-1} {V_{21}}^{\dagger}
(1-P_2) W_1P_1 V_1 \\
&+&
{V_1}^{\dagger} (1-P_1) ({V_{21}}^{\dagger} P_2
{W_2}^{\dagger})^{k} {V_{21}}^{\dagger}
(1-P_2) V_{21} (1-P_1) V_1\\
&+&
{V_1}^{\dagger} (1-P_1) ({V_{21}}^{\dagger} P_2
{W_2}^{\dagger})^{k}
{V_{21}}^{\dagger}
P_2 V_{21} (1-P_1) V_1
\ean
Two last terms here can be summed up to give a more simple
expression:
\ba\label{3}
\lefteqn{
{U_{k-1}}^{\dagger}U_{00} +
\sum_{n=0}^{\infty}{U_{n+k}}^{\dagger}U_n} \nonumber\\
&=&
{V_1}^{\dagger} (1-P_1) ({V_{21}}^{\dagger}
P_2 {W_2}^{\dagger})^{k-1} {V_{21}}^{\dagger}
(1-P_2) W_1P_1 V_1 \nonumber\\
&+&
{V_1}^{\dagger} (1-P_1) ({V_{21}}^{\dagger}
P_2 {W_2}^{\dagger})^{k} (1-P_1) V_1
\ea

In the r.h.s. of the latter expression however both terms are
equal to
zero. The reason is the following equations:
\be\label{2}
(1-P_2) W_1=0, \quad (1-P_1) W_2=0.
\ee
The first of them is a consequence of the fact that after
propagation in the wormhole $W_1$ the particle is in the space
region $S_2$. The operator $P_2$ projects
on this region, and $1-P_2$ projects on the complementary region
(all the space-like surface $t_2=\mbox{const}$ of the
background spacetime excluding the
region $S_2$). Going out of the wormhole $W_1$ the particle
cannot be in this complementary region. Therefore, the first
equation in (\ref{2}) is valid. Analogously, the second equation
in (\ref{2}) follows from the fact that after travelling in the
wormhole $W_2$ the particle turns out to be in the space region
$S_1$, and projecting, by $1-P_1$, on the complementary region
gives zero.

Using the first equation from (\ref{2}) and a conjugate to the
second equation we see that the expression (\ref{3}) as well as
its conjugate are equal to zero. Therefore,
\be\label{off-diag0}
{U_{k-1}}^{\dagger}U_{00} +
\sum_{n=0}^{\infty}{U_{n+k}}^{\dagger}U_n
=
{U_{00}}^{\dagger}U_{k-1} +
\sum_{n=0}^{\infty}{U_n}^{\dagger}U_{n+k}
= 0.
\ee
Thus, the off-diagonal terms do not contribute the sum
(\ref{tm-unitar-Thorne}), and both unitarity and generalized
unitarity are valid in the considered case: evolution from the
initial chronal region (earlier than the time machine) to the
final chronal region (after it).

This does not mean of course that both descriptions, one with the
superselection and one without superselection, are equivalent.
According to the coherent description, the evolution law is
following:
$$
\rho'=U\,\rho\, U^{\dagger}
= U_{00}\,\rho\, {U_{00}}^{\dagger}
+ \sum_{n=0}^{\infty}U_{00}\,\rho\, {U_n}^{\dagger}
+ \sum_{n=0}^{\infty}U_n\,\rho\, {U_{00}}^{\dagger}
+ \sum_{m,n=0}^{\infty}U_m\,\rho\, {U_n}^{\dagger}.
$$
According to the theory with decoherence (superselection), the
evolution should be described by partial evolution operators
according to a general formula (\ref{non-select-evolut}) that
takes the following form in our case:
\be\label{non-unitar}
\rho'=
U_{00}\,\rho \,{U_{00}}^{\dagger}
+\sum_{n=0}^{\infty}U_n\,\rho \,{U_n}^{\dagger}.
\ee
This leads of course to different predictions. Thus, the two
theories have different physical contents.\footnote{Let us
repeat once more that decoherence of different $n$ leading
to generalized unitarity must take place ``in ordinary
conditions'' while coherent description may be valid
``in special conditions'' when the environment does not
distinguish between different numbers of particles in the
dischronal region.}

Consider different locations of times $t'$, $t''$ (the start and
the end of the evolution) in respect to the time machine. In all
cases when at least one of the time moments $t'$, $t''$ is within
the time machine, no unitarity is found in the paper \cite{Th94}.
However, using the definition (\ref{in-fut-time-Q_2}), we can
prove Eq.~(\ref{off-diag0}) in just the same way as above for the
case $t' < t_1 < t'' < t_2$. Thus, in this case operator
(\ref{sum-U_n}) is also unitary. There is again no contradiction
with the result of \cite{Th94} because in that paper not this
operator, but $\sum_n K_n$ was considered as an evolution
operator (both definitions coincide for $t'<t_1<t_2<t''$).

Formally unitarity for $t' < t_1 < t'' < t_2$ is caused by the
projector $1-Q_2$ in the definition of $U_n$. Physically the
reason of this is that this projector provides escaping the
particle from the time machine (see Sect.~\ref{unitar-2}). Thus,
though we deal with $t'' < t_2$, actually the situation is, in a
sense, equivalent to the case $t''>t_2$.

For the cases $t_1<t'<t_2<t''$ and $t_1<t'<t''<t_2$ the relation
$U^{\dagger}U=1$ is invalid, and we are left only with the
generalized unitarity. We could introduce in these cases too a
sort of projector providing unitarity of the operator $\sum_n
U_n$. However
there is no physical ground for this (see the remark at the
end of Sect.~\ref{unitar-2}).

\section{States Trapped in the Time Machine}\label{trapped}

We considered the evolution of a particle entering the time
machine from the past chronal region, circulating several times
within the time machine and then escaping from it into the future
chronal region. There are however the states that exist only
within the time machine but not in the chronal regions. These
states may be described at arbitrary time moment between $t_1$
and $t_2$. We shall describe them at time $t_1$.

Consider the state $|\psi_1\rangle$ at time $t_1$ possessing the
properties
$$
(1-P_2) V_{21} |\psi_1\rangle  = 0, \quad
(1- W_2 V_{21}) |\psi_1\rangle  = 0
$$
It is evident that the particle in
this state ``bites its own tail'': after
evolution to $t_2$ it enters the wormhole $W_2$ and after going
out of the wormhole it turns out to be in the state
identical with the initial
state. After this the same cycle of evolution is repeated. The
particle in such a state travels through the interval $[t_1,t_2]$,
but it never passes any time moment before $t_1$ or
any time moment after $t_2$.

There is evident generalization of this state, providing passing
the dischronal region twice, three times or generally $n$
times before reaching the initial state.
Such a state satisfies the following
conditions at time $t_1$:
\ba\label{n-trap}
(1-P_2) \,(V_{21} W_2)^{n'} V_{21} |\psi_n\rangle
&=& 0 \mbox{ for }n'<n,\nonumber\\
(1- (W_2 V_{21})^n) |\psi_n\rangle  &=& 0.
\ea
Then the particle ``bites its own tail'' after
winding $n$ causal loops:
after $n$ cycles, each of which contains travelling through the
interval $[t_1,t_2]$, entering the wormhole $W_2$ and passing
this wormhole, the state is identical to the initial one.

The states thus described are nothing else than a quantum
version of
the classical ``Jinnee'' discussed in \cite{jinn1,jinn2}, the
classical bodies moving along closed time-like trajectories
within a time machine.

One can construct a state in such a way as to provide passing the
dischronal region infinite number of times never escaping to
the future, but with the initial state never being repeated.
The condition for this is (at time $t_1$)
\be\label{inf-trap}
(1-P_2) \,(V_{21} W_2)^{n'} V_{21} |\psi_{\infty}\rangle  = 0
\mbox{ (all }n').
\ee
The particle in such a state cannot go
out of the time machine though probably its state never becomes
identical to the initial state. The states (\ref{n-trap}) may
be considered to be special cases of (\ref{inf-trap}).

One more generalization can be considered: the state existing
earlier than the time machine emerged (before $t_1$) but trapped
in TM so that it cannot be found at times later than $t_2$. Let
us characterize this state at time $t'<t_1$:
\ban
P_1 V_1 |\psi_n\rangle  &=& 0
        \mbox{ (the particle does not enter $W_1$)},\\
(1-P_2) (V_{21}W_2)^{n'} V_{21} (1-P_1) V_1 |\psi_n\rangle  &=& 0,
\quad n'\ge 1
        \mbox{ (it does not escape TM)}.
\ean
It may be shown that the state of the particle in this case cannot
be repeated after a finite number $n$ of cycles. The condition
for such a repetition,
$$
(1 - (V_{21}W_2)^n) V_{21} (1-P_1) V_1 |\psi_n\rangle = 0
        \mbox{ (the state is repeated after $n$ cycles)}.
$$
would be inconsistent due to the relation
$
W_2  = P_1 W_2
$
(expressing that the mouth leading out from the wormhole $W_1$ is
in the region $S_1$).

Therefore, it is possible that a
particle enters the time machine from the past and stays within
it, infinitely repeating evolution from $t_1$ to $t_2$ and
backward. Moreover, the states
obtained by time reversal from those already considered,
may also be defined. Then the particle escapes
from the time machine into the future after infinite number of
cycles within the time machine, never being in the past chronal
region.

Existence of the states trapped in the time machine contradicts
to the generalized unitarity condition in the form given earlier.
Indeed, this condition provides that any state evolves finally
to the
future chronal region through one of the channels enumerated by
$n$. The trapped states do not at all escape from the time
machine. These states belong to the class of `finely
tuned' states which the generalized unitarity condition is not
valid for (see Remark~\ref{fine-tune} at
page~\pageref{fine-tune}).

For example, application of the (generalized) unitarity condition
(\ref{both-in-unitar}) for evolution from the time $t_1$
(considered as a time within the time machine) to the same time
$t_1$ should seemingly forbid the trapped states. However,
analysing attentively the proof of this condition, we see that
it is supposed in this proof that the operator
$$
I_N = 1-\sum_{n=0}^{N} {U_n}^{\dagger}\,U_n
= \left({V_{21}}^{\dagger}\,P_2\,{W_2}^{\dagger}\right)^N
{V_{21}}^{\dagger}\,P_2\, V_{21}
\left(W_2 \,P_2\, V_{21}\right)^N
$$
tends to zero when $N\to\infty$. The expectation value of
this operator for almost
all states (as well as the trace of this operator multiplied by
almost all density matrices) decreases with increasing $N$.
However, for the states considered in the present section (the
trapped states) we have, as a consequence of
Eq.~(\ref{inf-trap}),
$$
\langle\psi | \,
1-\sum_{n=0}^{N} {U_n}^{\dagger}\,U_n \,
| \psi\rangle = \langle\psi | \psi\rangle = 1.
$$
Hence the operator $I_N$ does not decrease and the generalized
unitarity condition is not fulfilled in the space of the trapped
states.

The physical interpretation of this fact in terms of
probabilities is following. For this very special class of
states (not escaping to the future chronal region) no part of
probability goes into channels enumerated by $n$. It might be
reasonable to introduce one more channel corresponding to
$n=\infty$ and claim that all probability, for these states,
goes into this channel. This leads to the following remark.

\begin{remark} {\rm
In a general case (for arbitrary states) the generalized
unitarity condition should be written in the form
$$
{U_{00}}^{\dagger}U_{00} + \sum_{n=0}^{N} {U_n}^{\dagger}U_n
+ I_N = {\bf 1}
$$
(where the operator $I_N$ is given above for the case
$t'=t''=t_1$ and can be readily written for all other choices of
$t'$ and $t''$). The expectation value of this formula for an
arbitrary state $|\psi\rangle$,
$$
\langle\psi |{U_{00}}^{\dagger}U_{00} |\psi\rangle
+ \sum_{n=0}^{N} \langle\psi | {U_n}^{\dagger}U_n |\psi\rangle
+ \langle\psi | I_N |\psi\rangle = 1,
$$
gives the probability distribution for different channels of
evolution (the last term in the l.h.s. gives the probability for
all channels with numbers more than $N$). If the limit
$$
\langle\psi |I_{\infty}|\psi\rangle
=\lim_{N\to \infty}\langle\psi |I_N |\psi\rangle
$$
exists, the formula
$$
\langle\psi |{U_{00}}^{\dagger}U_{00} |\psi\rangle
+ \sum_{n=0}^{\infty} \langle\psi | {U_n}^{\dagger}U_n |\psi\rangle
+ \langle\psi |I_{\infty}|\psi\rangle = 1
$$
gives the probability distribution for all channels with finite
$n$ and for the channel with $n=\infty$ corresponding to the
trapped states.
}\end{remark}

\section{Conclusion}\label{conclus}

We considered in this paper how evolution of a non-relativistic
non-interacting quantum particle in the spacetime with closed
time-like curves (a time-machine spacetime) should be described.
A simple non-relativistic model of such a spacetime was used
corresponding to emergence of the time machine at some time
moment and its disappearance at another moment.

In the paper \cite{Th94} evolution of a particle in this
spacetime was described with the help of the evolution operator
$U$ found by path integration. It was shown that such an operator
is multiplicative and unitary only when the evolution is
considered between the time moments belonging
to the chronal regions (including
the case when one of the time moments is in the past chronal
region and the other is in the future chronal region).

In our paper the arguments were given that superselection
(decoherence) may arise for the evolution in this spacetime
(in ``normal conditions''). The superselection sectors are
enumerated in this case by the number of times
the particle returns to its past. Therefore, evolution of the
particle is described by a family of {\em partial evolution
operators} $U_n$ (or partial propagators) instead of a single
evolution operator $U$.

In the case when the evolution is considered into the future
chronal region (i.e. to the time moment after the time machine
disappearance), the partial propagators may be correctly defined
as integrals over the corresponding topological classes of paths.
If the final time of evolution is within the time machine, the
path-integral expressions for the partial propagators (evolution
operators) should be corrected by the projectors providing
escaping the particle from the time machine after a certain
number of returns to the past.

With these definitions, the family of partial evolution operators
satisfy the generalized unitarity condition $\sum_n
{U_n}^{\dagger}U_n=1$ and multiplicativity condition $U_n
U_m=U_{n+m}$ (the latter is fulfilled only
for at least one of the numbers $n$, $m$ equal to zero). It is
shown that the generalized unitarity is compatible with the
unitarity of the operator $U=\sum_n U_n$ in the case of evolution
from the past of the time machine to its future. Nevertheless, the
operator $U$ cannot describe correctly the evolution if the
superselection exists.

A special class of states of the particle is described emerging
and disappearing simultaneously with the time machine, i.e.
existing only within it. These states present a quantum analogue
of the classical ``Jinnee'' states investigated earlier
\cite{jinn1,jinn2}. Besides this, the states are considered that
either 1)~exist in the past of the time machine but then are
trapped in it and never escape into the future of the time
machine, or 2)~never exist in the past of the time machine, have
infinite number of cycles within the time machine and then
finally escape into its future. For all these states the
generalized unitarity condition should be corrected to take into
account the channel of evolution with $n=\infty$.

From conceptual point of view, the superselection of the
considered type may be thought of as the influence of the future
on the past, i.e. a sort of ``consistency conditions''.

The coherent (unitary) evolution in the time-machine spacetime
is also possible, but only in quite special conditions when the
environment of the particle in the dischronal region does not
distinguish between different numbers of particles.\footnote{This
question needs special investigation with a concrete
model of an
environment. A very good discussion of the role of environment
in different situations may be found in \cite{Zeh95}.} It is
interesting that in the case of coherent evolution no
information from the future to the past can be carried by the
particle. If the particle leaves some information about the
future in its past, then the interaction responsible for
the information transfer leads to decoherence so that the
winding number $n$ becomes definite.

\vspace{0.5cm}
\centerline{\bf ACKNOWLEDGEMENTS}

One of the authors (M.M.) is grateful to J.\ Audretsch,
V.A.\ Namiot and H.-D.\ Zeh for discussing the question
of decoherence. The work was supported in part by the
Danish National Research Foundation through the
establishment of the Theoretical Astrophysics Center,
the Danish Natural Science Research Council through
grant 11-9640-1, and by the Deutsche Forschungsgemeinschaft.


\begin{thebibliography}{10}

\bibitem{Th91} K.S.Thorne, Ann. New York Acad. Sci. {\bf 631},
182 (1991).
\bibitem{Haw92} S.W.Hawking, Phys. Rev. {\bf D~46}, 603 (1992).
\bibitem{Th93} K.S.Thorne, {\em Proceedings of the 13th
International Conference on General Relativity and Gravitation},
ed. C.Kozameh, IOP Publishing: Bristol (1993).

\bibitem{Fr90} J.L.Friedman, M.S.Morris, I.D.Novikov,
F.Echeverria, G.Klink\-ham\-mer, K.S.Thorne, and U.Yurtsever,
Phys. Rev. {\bf D~42}, 1915 (1990).

\bibitem{Fr91} J.L.Friedman and M.S.Morris, Phys. Rev. Lett.
{\bf 66}, 401 (1991).

\bibitem{Ech91} F.Echeverria, G.Klinkhammer, and K.S.Thorne,
Phys. Rev. {\bf D~44}, 1077 (1991).

\bibitem{Nov92} I.D.Novikov, Phys. Rev. {\bf D~45}, 1989 (1992);
A.Lossev and I.D.Novikov, Class. Quant. Grav. {\bf 9}, 2309
(1992).

\bibitem{Nov93} E.V.Mikheeva and I.D.Novikov, Phys. Rev. {\bf
D~47}, 1432 (1993).

\bibitem{Pol92} H.D.Politzer, Phys. Rev. {\bf D~46}, 4470 (1992).

\bibitem{Fr92} J.L.Friedman, N.J.Papastamatiou, and J.Z.Simon,
Phys. Rev. {\bf D~46}, 4442 (1992); {\bf D~46}, 4456 (1992).

\bibitem{Th94}D.S.Goldwirth, M.J.Perry, T. Piran, and K.S.Thorne,
Phys. Rev. {\bf D~49}, 3951 (1994).

\bibitem{Zeh73} H.D.Zeh,
Toward a quantum theory of observation,
Found. Phys. {\bf 3}, 109 (1973).

\bibitem{Zurek81} W.H.Zurek, Phys. Rev. {\bf D~24}, 1516 (1981).

\bibitem{Zurek82} W.H.Zurek, Phys. Rev. {\bf D~26}, 1862 (1982).

\bibitem{JoosZeh85} E.Joos and H.D.Zeh,
Z.Phys. {\bf B~29}, 223 (1985).

\bibitem{Men79a} M.B.Mensky, Phys. Rev. {\bf D~20}, 384 (1979).

\bibitem{Men79b} M.B.Mensky, Sov. Phys.-JETP {\bf 50}, 667
(1979).

\bibitem{Men-book3}M.B.Mensky, {\em Continuous Quantum
Measurements and Path Integrals}, IOP Publishing: Bristol and
Philadelphia, 1993.

\bibitem{jinn1} A.Lossev, I.Novikov, Class. Quantum Grav. {\bf
9}, 1 (1992).

\bibitem{jinn2} I.Novikov, in: {\em String Quantum Gravity and
Physics at the Planck Energy Scale} (Proceed. Intern. Workshop on
Theoretical Physics, Erice, Italy, 21-28 June 1992), edited by
N.Sanchez, Singapore: World Scientific, 1993, p.480.

\bibitem{Zeh95} D.Giulini, C.Kiefer, H.D.Zeh,
Phys. Lett. {\bf A~199}, 291 (1995).

\end{thebibliography}
\end{document}